\documentclass[a4paper,fleqn,usenatbib]{mnras}
\usepackage[T1]{fontenc}
\usepackage{ae,aecompl}
\usepackage{graphicx}
\usepackage{amsmath}
\usepackage{amssymb}
\usepackage{ulem}
\title[Large disc of IRAS 22150+6109]
{IRAS\,22150+6109 -- a young B-type star with a large disc}

\author[O. V. Zakhozhay et al.]{
Olga V. Zakhozhay$^{1,2,3,4}$\thanks{E-mail:zkho@mao.kiev.ua}
Anatoly S. Miroshnichenko$^{2,3,5}$, Kenesken S. Kuratov$^{2,3}$, 
\newauthor Vladimir A. Zakhozhay$^{6,7}$, Serik A. Khokhlov$^{2,5,7}$, Sergey V. Zharikov$^8$,
\newauthor Nadine Manset$^9$
\\
$^1$Main Astronomical Observatory, National Academy of Sciences of Ukraine, Kyiv 03680, Ukraine\\
$^2$Fesenkov Astrophysical Institute, Observatory, 23, Almaty, 050023, Kazakhstan\\
$^3$NNLOT, Al-Farabi Kazakh National University, Almaty 050040, Kazakhstan\\
$^4$Max Planck Institute for Astronomy, K\textbf{\"o}nigstuhl 17 D-69117 Heidelberg, Germany\\
$^5$Department of Physics and Astronomy, University of North Carolina at Greensboro, Greensboro, NC 27402--6170, USA\\
$^6$V.~N.~Karazin Kharkiv National University, Kharkiv, Ukraine\\
$^7$Faculty of Physics and Technology, Al-Farabi Kazakh National University, Almaty 050040, Kazakhstan\\
$^8$Instituto de Astronom\'{\i}a, Universidad Nacional Aut\'{o}noma de M\'exico, Ensenada, Baja California, 22800, Mexico\\
$^9$CFHT Corporation, 65--1238 Mamalahoa Hwy, Kamuela, HI 96743, USA
}

\date{Accepted XXX. Received YYY; in original form ZZZ}

\pubyear{2017}

\begin{document}
\label{firstpage}
\pagerange{\pageref{firstpage}--\pageref{lastpage}}
\maketitle

\begin{abstract}
We present the results of a spectroscopic analysis and spectral energy distribution (SED) modelling of the optical counterpart of the infrared source IRAS\,22150+6109. The source was suggested to be as a Herbig~Be star located in the star forming region L\,1188. Absorption lines in the optical spectrum indicate a spectral type B3, while weak Balmer emission lines reflect the presence of a circumstellar gaseous disc. The star shows no excess radiation in the near-infrared spectral region and a strong excess in the far-infrared that we interpret as radiation from a large disc, whose inner edge is located very far from the star (550~au) and does not attenuate its radiation.  We conclude that IRAS\,22150+6109 is an intermediate-mass star that is currently undergoing a short pre-main-sequence evolutionary stage.
\end{abstract}


\section{Introduction}
\label{introduction}

IRAS~22150+6109 is an infrared source located in the direction of an active star-forming region L~1188 in Cepheus \citep{Abraham95}.
It was identified with a poorly studied  $V \sim$ 11 mag star listed in the catalogue of early-type emission-line objects by \citet{Wackerling70}. It is also was detected by the Hamburg survey for emission-line stars \citep{Kohoutek99} and included in a catalogue of reflection nebulae \citep{Magakian03}.  The object exhibits a strong infrared excess that can be attributed to the presence of an intermediate-mass star in transition from the pre-main-sequence to the main-sequence stage of evolution. A weak H$\alpha$ emission in the IRAS~22150+6109 spectrum also supports the nearly main-sequence status of the source. Results on the object's CO emission are controversial: a negative detection by \citet{Wouterloot89} and a positive detection by \citet{Kerton03}. No emission from H$_2$O, OH, or CS molecules has been detected from its direction \citep{Wouterloot93, Bronfman96}. 

In this paper we present new high-resolution optical spectroscopy of the IRAS~22150+6109 obtained in 2004--2016 together with analysis of the optical photometry, which was taken in 1997--1999.   We  additionally collected all currently available infrared photometric data  
and included them in our modelling of the object's spectral energy distribution (SED) with the aim to probe the properties of the circumstellar material. 

 The paper is organized as follows. Section 2 presents the description and brief analysis of the spectroscopic and photometric observations. The modelling approach and the basic equations for the SEDs modelling are briefly described in Section 3. The best-fit results for the disc and the discussion are presented in Section 4. Section 5 includes the conclusions of the paper.

\section{Observations}
\subsection{Spectroscopy}
\label{spectra}
Four high-resolution spectroscopic observations of IRAS~22150+6109 were taken in 2004 at the 3.6~m Canada--France--Hawaii Telescope (CFHT) and in 2015/2016 at the 2.1~m telescope of the Observatorio Astron\'omico Nacional San Pedro Martir (OAN SPM). The OAN SPM data were reduced in a standard way with the $echelle/slit$ package in IRAF.\footnote{IRAF is distributed by the National Optical Astronomy Observatory, which is operated by the Association of Universities for Research in Astronomy (AURA) under a cooperative agreement with the National Science Foundation.} Observations obtained with the ESPaDoNS spectropolarimeter in the spectroscopic mode at CFHT were reduced with Upena and Libre-ESpRIT software packages \citep{1997MNRAS.291..658D}. The observing dates, facilities, resolving power and spectral ranges are listed in Table~\ref{tabl_1}.

\begin{figure}
\centering
 \includegraphics[width=8.0cm, bb =  40   10   710   520, clip=]{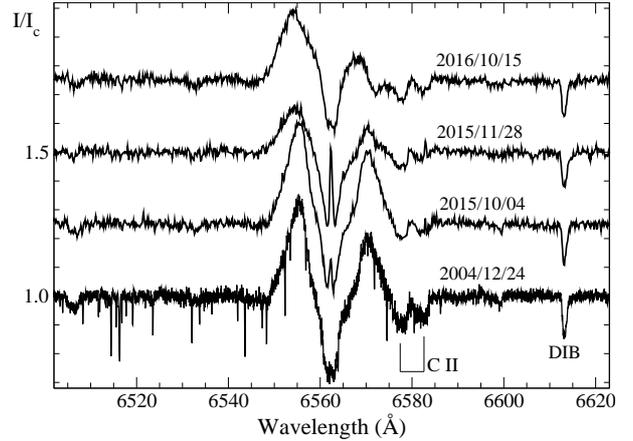} 
 \caption{The H$\alpha$ line profile variations in the spectra of IRAS~22150+6109. The intensity is normalized to the local continuum (I$_{\rm c}$), the wavelengths scale is heliocentric. The narrow absorption lines in the 2004 spectrum are telluric. The spectra are vertically shifted with respect to each other by 0.25\,I$_{\rm c}$.}
   \label{Halpha_var}  
\end{figure}

There are several emission features in the spectrum. The strongest one is a double-peaked H$\alpha$ line (Fig.~\ref{Halpha_var}). The emission peaks are separated by $\sim 680\pm30$ km\,s$^{-1}$. They move noticeably beyond the measurement errors ($\pm 10$ km\,s$^{-1}$). The largest blueward shift by $\sim$ 150 km\,s$^{-1}$ was detected in our 2016 spectrum. The line also exhibits a variable narrow central emission peak at a systemic velocity of $-20\pm4$ km\,s$^{-1}$ determined by averaging positions of 10 mostly He~{\sc i} absorption lines. This peak is almost invisible in 2004 and 2016 and much stronger in 2015. All these features may be due to a variable accretion onto the young star.

\begin{figure}
\centering
 \includegraphics[width=8.0cm, bb =  40   10   710   520, clip=]{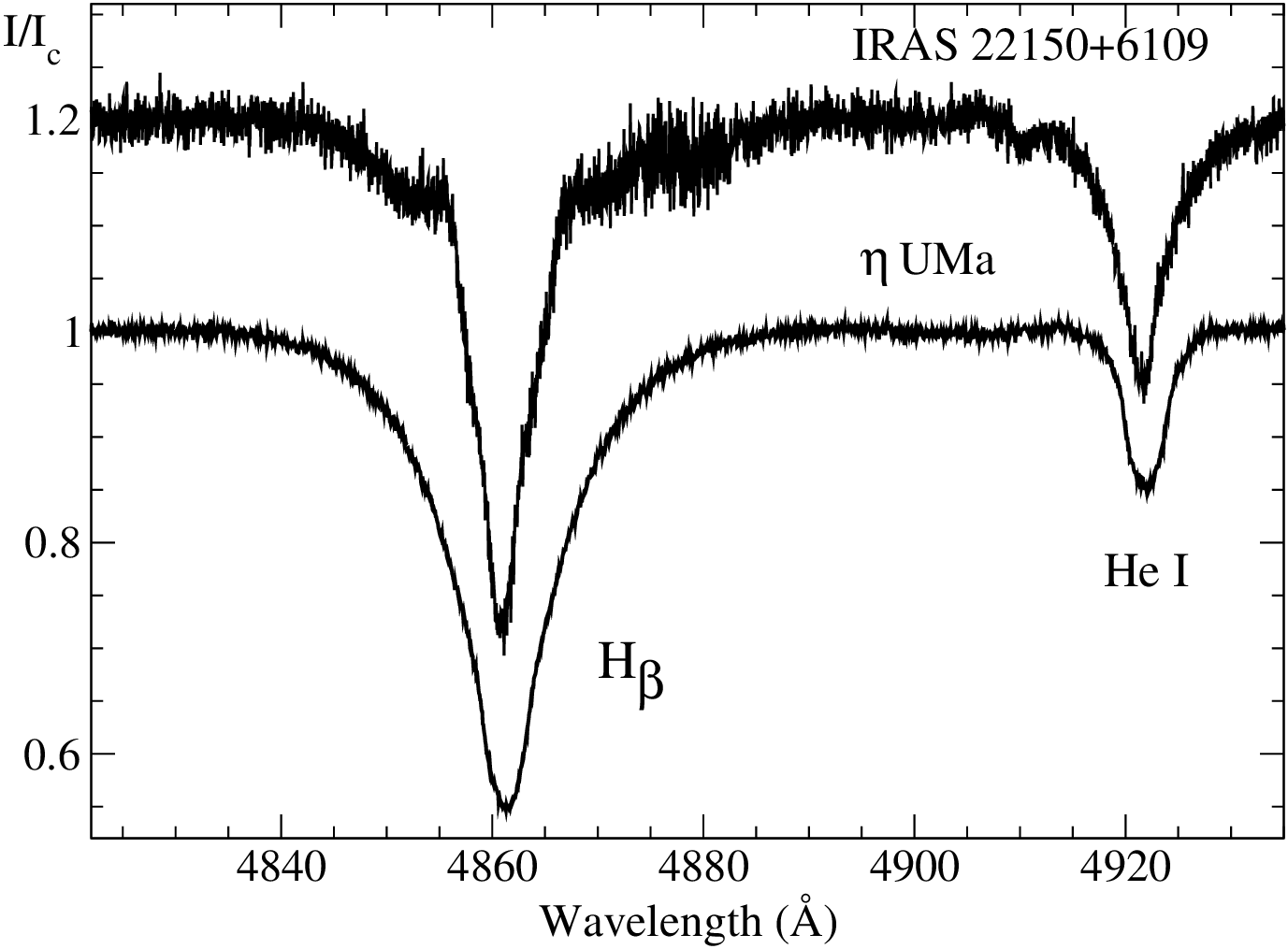} 
 \caption{Comparison of the H$\beta$ line region in the CFHT spectrum of IRAS~22150+6109 with that of $\eta$ UMa. The intensity and wavelength are plotted in the same way as in Fig.\,\ref{Halpha_var}.}
   \label{Hb}  
\end{figure}

\begin{figure}
\centering
 \includegraphics[width=8.0cm, bb =  40   20   710   520, clip=]{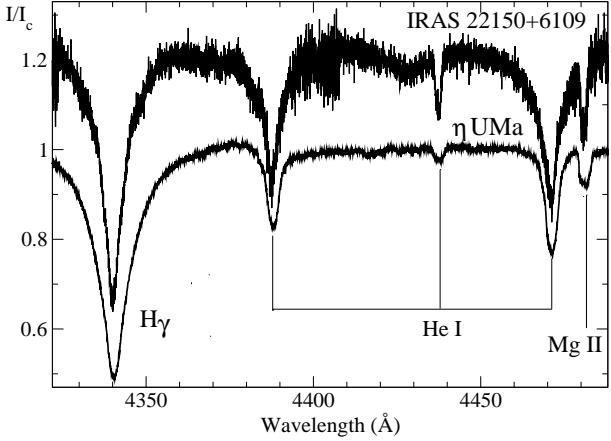} 
 \caption{Comparison of the H$\gamma$ line region in the CFHT spectrum of IRAS~22150+6109 with that of $\eta$ UMa. The intensity and wavelength are plotted in the same way as in Fig.\,\ref{Halpha_var}.}
   \label{Hg}  
\end{figure}

The absorption-line spectrum shows moderately broad lines that are consistent with a spectral type B3 and a projected rotational
velocity of v\,$\sin i \sim$200 km\,s$^{-1}$ (Fig.~\ref{Hb} and Fig.~\ref{Hg}). The latter was estimated by comparison with a spectrum of $\eta$ UMa (B3{\sc v}, v\,$\sin i \sim$150 km\,s$^{-1}$) taken from the archive of the spectrograph ELODIE \citep{Moultaka04}.

Relatively large projectional rotation velocity indicates that the star's rotation axis (and the disc, as a consequence) is significantly tilted from the line of sight. Assuming the star's mass is 7~M$_{\odot}$ and radius 3.3~R$_{\odot}$, the critical rotation velocity is equal to v$_{\rm crit}$ = 436 $\times$ (M/M$_{\odot}$) / (R/R$_{\odot}$) = 925 km\,s$^{-1}$. The ratio of the observed projected rotation velocity to the critical velocity sets a limit on the tilt angle as $i > \arcsin(200/925) = 13\degr$.

Other emission features detected are very weak forbidden oxygen lines [O~{\sc i}] 6300 and 6364 \AA\ and hydrogen lines of the Paschen series. No noticeable line profile variations, except for the mentioned above, were found in our spectra.

\begin{figure}
\centering
 \includegraphics[width=8.0cm, bb =  40   10   710   520, clip=]{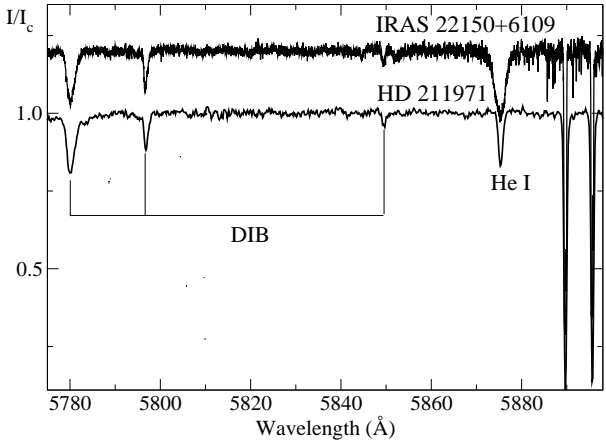} 
 \caption{Comparison of DIBs and interstellar Na lines in our CFHT spectrum of IRAS~22150+6109 with those in the spectrum of HD\,211971 taken at the Three College Observatory (North Carolina, USA, resolving power $R \sim$12000). The intensity and wavelength are plotted in the same way as in Fig.\,\ref{Halpha_var}.}
   \label{DIB}  
\end{figure}

Interstellar features are represented by diffuse interstellar bands (DIBs) and absorption lines of Na {\sc i} (D--lines at 5889 and 5895~\AA) and K~{\sc i} (7699 \AA). The DIBs and interstellar lines strength is consistent with an optical reddening of E$(B-V) = 0.7$ mag. Both the DIBs and Na~{\sc i}~D--lines in the spectrum of IRAS~22150+6109 are very similar to those of HD\,211971(Fig.~\ref{DIB}), an somewhat more reddened (E$(B-V) = 0.9$ mag) A2~{\sc i}b supergiant located in 1$\fdg$5 from the object. Recent GAIA measurements give a distance of $862^{+237}_{-153}$~pc for IRAS\,22150+6109 and $1149^{+428}_{-215}$~pc for HD\,211971, respectively \citep{GAIA}. This comparison shows that IRAS\,22150+6109 is not affected by a noticeable circumstellar reddening.

\begin{table*}
\centering
\caption{The summary of the observing dates and instruments.}
\label{tabl_1}
\begin{tabular}{cclcccc}
\hline
Date       & MJD &Telescope      & Location                & Resolv.power & Sp.range, \AA\\
\hline
2004/12/24 & 3364.765 & 3.6\,m CFHT        & Hawaii, USA                  & 65000    & 4000--10500  \\
2015/10/04 & 7300.648 & 2.1\,m OAN SPM & Baja California, Mexico & 18000    & 3650--7320   \\
2015/11/28 & 7355.584 & 2.1\,m OAN SPM & Baja California, Mexico & 18000    & 3650--7320   \\
2016/10/15 & 7677.714 & 2.1\,m OAN SPM & Baja California, Mexico & 18000    & 3650--7320   \\
\hline
\end{tabular}
\end{table*}

\subsection{Photometry}
\label{Photometric_data}
 The multicolor $UBVRI$ photometry of IRAS\,22150+6109 was obtained at the Tien-Shan Astronomical Observatory, Kazakhstan using the FP3U  photometer attached to a 1\,m telescope \citep{Bergner88}. The results of 10 observations obtained in 1997--1999 were published by \citet{Kuratov04}. The infrared photometric data for the object were extracted from 2MASS, WISE, and AKARI databases \citep[][, respectively]{Cutri03,Wright10,Murakami07}. The magnitudes were converted to fluxes using zero-points from  \citet{Straizhys77} and descriptions of the mentioned catalogues. 

The object's optical brightness varies with an amplitude of 0.2~mag. The average $V$--band brightness is 10.82$\pm$0.07 mag. Similar variations with an amplitude of 0.15 mag were detected by the NSVS survey in 1999--2000 \citep[no filter optical photometry,][]{Wozniak2004}. The average color-indices ($U-B = -0.37\pm0.05$, $B-V = 0.46\pm0.03$, $V-R = 0.40\pm0.05$, $R-I = 0.28\pm0.03$) suggest a spectral type of B3 and an extinction of A$_{\rm V} = 2.0\pm$0.1~mag. 
The SED of the object is shown in  Fig.~\ref{best}.
The observed SED was dereddened using a standard Galactic interstellar extinction law from \citet{Savage79}.

\section{Modelling of spectral energy distribution}\label{model}

\subsection{Modelling algorithm}

We consider the SED of IRAS~22150+6109 to be composed of a B3 {\sc ZAMS}-star and a protoplanetary disc. We model the stellar fluxes in a blackbody approximation

\begin{equation}
f_{\nu,\ast} = d^{-2} \pi R_{\ast}^{2} B_{\nu} (T_{\ast}),
\end{equation}
where $d$ is the distance to the star, $B_{\nu} (T)$ is the Plank function, $R_{\ast}$ and $T_{\ast}$ are the stellar radius and effective temperature. This equation does not take into account extinction of the stellar radiation in the disc, because the optical photometric and spectroscopic data show no evidence for the presence of circumstellar extinction. For the computations we use $R_{\ast} = 3.3 R_{\odot}$ that we derived from the observed visual magnitude, interstellar reddening, $T_{\ast} = 20 000$~K, and bolometric correction \citep{mir97}. The stellar parameters are consistent with the data of \cite{SK1981} and the mentioned above distance estimates within their uncertainty range. 
 
It was assumed that the entire excess radiation at wavelengths $\lambda > 1 \mu$m originates from a protoplanetary disc.
The flux density in this case is given by

\begin{equation}\label{form7}
 f_{\nu,disc}= d^{-2} \int_{R_{in}}^{R_{out}}B_{\nu}(T_r) (1 - exp(-\tau_{\nu,r})) 2\pi r dr,  
 \end{equation}
where $R_{in}$ and $R_{out}$ are the disc inner and outer radii, respectively, and $\tau_{\nu,r}$  is the optical depth of the disc material. The latter is the product of a wavelength-dependent disc opacity, $\kappa_{\nu}$, and the radial surface density distribution, $\varSigma_r$. To calculate the disc opacity, we used the Mie theory and considered spherical grains composed of astronomical silicates, a density of 2.5~g\,cm$^{-3}$, and sizes between 0.01 and 100~$\mu$m (see \citealt{Boehler13} for a detailed description of the grain emissivity determination). We assume that the disc is young and has a gas to dust mass ratio of 100.

Since the uncertainty of the distance to the star is quite large (see Section\,\ref{spectra}), we normalize all the fluxes to its $V$--band flux 
corrected for the interstellar extinction (see Section\,\ref{Photometric_data}).

\par The temperature $T_r$, vertical height $H_r$ and surface density $\varSigma_r$ distributions are taken to be power laws in disc radius:
\begin{equation}\label{form10}
  T_r=T_{in}\left(\frac{r}{R_{in}}\right)^{-q},
    \end{equation}
    
 \begin{equation}\label{Hr}
  H_r=H_{in}\left(\frac{r}{R_{in}}\right)^{\frac{3-q}{2}},
    \end{equation}   
    
\begin{equation}\label{form9}
  \varSigma_r=\varSigma_{in}\left(\frac{r}{R_{in}}\right)^{-p},
    \end{equation}
where $T_{in}$ is the disc temperature at $R_{in}$, that we calculate using the radiative equilibrium equation, $R_{\ast}$, and $T_{\ast}$. The index $q$ and the disc vertical height at the inner edge $H_{in}$ are taken to be free parameters. The surface density index $p = 1.5$ we chose to match the mass surface density of the current Solar system \citep{Carpenter09}. $\varSigma_{in}$ is a surface density at the inner disc radius $R_{in}$. The total mass of the disc is related to the disc size and surface density as follows:
\begin{equation}\label{md}
M_d = \int_{R_{in}}^{R_{out}}2 \pi r~\varSigma_r~{\rm d}r.
\end{equation}
Conversely, one can express $\varSigma_{in}$ as a function of the disc mass:
\begin{equation}\label{Sigma_r}
\varSigma_{in} = \frac {M_d~R_{in}^{-p}~(2-p)}{2 \pi~(R_{out}^{2-p} - R_{in}^{2-p})}.
\end{equation}
The geometry of a disc tilted with respect to the line of sight is accounted for by the approach described in \citet{Zakhozhay15}. It assumes that the disc edges at the inner and outer radii have a cutoff flat geometry and emit as blackbodies with constant temperatures derived with the radiative equilibrium equation and the equation~\ref{form10} for the inner and outer edges, respectively.\\

\section{SED modelling results and discussion}

\par Using the modelling approach described in the previous section, a grid of SEDs for the disc with different parameters was calculated. The ranges and increments of the modelling parameters are shown in Table\,\ref{t2}.

\par
\begin{figure}
\centering
 \includegraphics[width=\columnwidth]{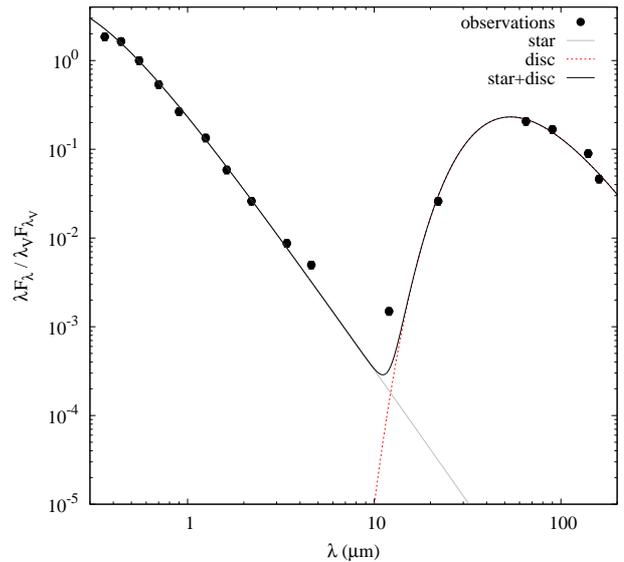} 
 \caption{The best-fit disc model (black solid line) is shown in comparison with the observed photometric SED of IRAS~22150+6109 (black dots). The  photospheric emission of B3~ZAMS star is shown with the gray solid line.}
   \label{best}  
\end{figure}

 The best fit of the SED we found by minimizing
 
\begin{equation}\label{formX2}
  \chi^2=\sum^{n}_{i=1}\Big({\frac{F_{obs,i}-F_{mod,i}}{F_{obs,i}}}\Big)^2,
    \end{equation}
where $F_{obs,i}$ and $F_{mod,i}$ are the observed and modelled fluxes (at the corresponding wavelength) respectively. The observational errors we assume to be equal to 10$\%$ of the corresponding $F_{obs,i}$ for all the observations. We normalize all the differences to the observed fluxes to account for the large range of the observed fluxes ($\sim 10^3$, see Fig.\,\ref{best}). We consider the disc model fits only at wavelengths $>$~1~$\mu$m assuming that the disc emission contributes significantly only in this region.

\begin{table}
\begin{minipage}{84 mm}
\centering
\caption{Modelling parameters}
\label{t2}
\begin{tabular}{llll}
\hline 
Parameter      & Best-fit                    & Range                               & Increment\\
\hline
$R_{\rm in}$  &  $550^{+70}_{-60}$~au         & 10 $\div$ 1000 au                           & 10 au     \\
$R_{\rm out}$ &  $700^{+100}_{-80}$~au      & 10 $\div$ 3000 au                         & 10 au   \\
$q$           &  $0.75^{+0}_{-0.25}$      & $0.35 \div 0.75$                        & 0.01     \\
$M_{\rm d}$   &  $0.2^{+1}_{-0.18}M_{\odot}^{\rm a}$  & $10^{-5} \div 7.0 M_{\odot}^{\rm b}$       & 10$\%$\\
$H_{\rm in}$  &  $H_{\rm in} < 0.08R_{\rm in}^{\rm c}$         & $10^{-5}\div 10^{-1}R_{\rm in}$                    & 10$\%$        \\
$i$           &  $>50\degr$              & $20\degr^{\rm d} \div 60\degr^{\rm e}$    & 10\degr\\
\hline
\end{tabular}
\end{minipage}

\begin{minipage}{84 mm}
\medskip
$^{\rm a}$ -- $=0.03M_{\ast}$, assuming that $M_{\ast} = 7M_{\odot}$.\\
$^{\rm b}$ -- the limit was chosen on the assumption that any putative disc should not be more massive than the central star.\\
$^{\rm c}$ -- the best fit has a disc with $H_{in} = 10^{-5} R_{in}$, although decrease of $H_{in}$ improves the fit insignificantly. \\
$^{\rm d}$ -- the lower limit determined by the spectral data: the ratio of the observed projected rotational velocity to the critical velocity indicates that the tilt angle of the star's rotational axis is $i >$ 13\degr.\\
$^{\rm e}$ -- the upper limit is based on the requirement that disc material should not block a part of the emission that comes from the star (for more details see \cite{Zakhozhay15}).
\end{minipage}
\end{table}
 
\par The best-fit ($\chi^2$ = 0.85) system parameters are listed in the second column of Table\,\ref{t2} along with their 1~$\sigma$ uncertainties calculated from the $\Delta\chi^2$ confidence statistics. The upper limit for $M_d$ reflects the assumption that the disc total mass should not exceed the mass of the central star. The actual upper limit is impossible determine without additional observations at longer wavelengths. Figure~\ref{best} shows the photometric data from observations (black filled circles), the assumed SED of the star (gray line), and  the resulting  SED of the star+disc system for the best fit parameters (black line). The errors of the observational points do not exceed the circle symbol size.
 
\subsection{Effect of reasonable variations of the free parameters}

\par
\begin{figure*}
\centering
 \includegraphics[width=12.0cm]{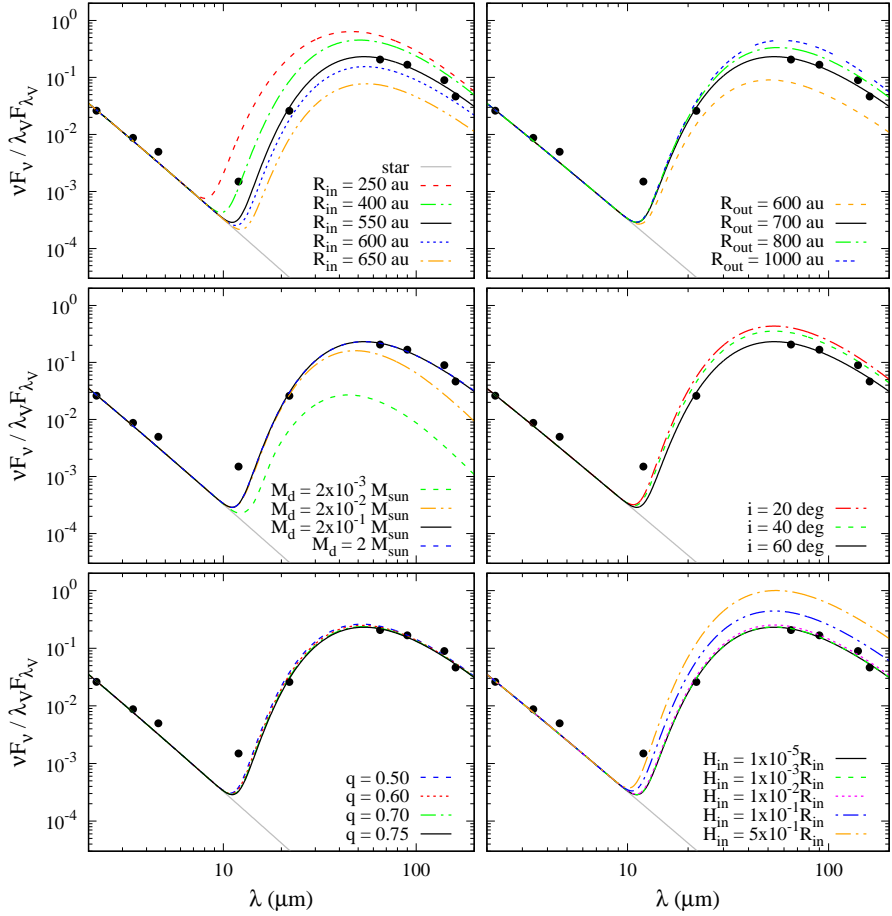} 
 \caption{ Effects of reasonable variations of the model parameters on the system SED. The black solid line in each panel marks the system SED for the best-fit disc parameters: $R_{in}$ = 550 au, $R_{out}$ = 700 au, $M_d = 0.2 M_{\odot}$, $i = 60^{\circ}$, $H_{in}=10^{-5}$$R_{in}$ and $q = 0.75$. The varied parameter from this fiducial set is indicated in each panel. The photospheric radiation of a B3~ZAMS star is shown with the gray solid lines.}
   \label{SEDs_var}  
\end{figure*}

 We have analysed how varying each of the 6 free parameters in our model affects the total SED from the system (see Fig.~\ref{SEDs_var}). The black solid line in each panel shows the total flux from the system with the best-fit disc model, the gray solid line shows the flux from the star. All the other lines show the models with all the best-fit parameters except for the varied one indicated in each panel.

\par One can see in Fig.~\ref{SEDs_var} that the two fit parameters determined with a highest confidence are $R_{in}$ and $R_{out}$. Their variations change the SED the most. A smaller $R_{in}$ = 250~au causes the maximum flux to increase 31.4 times at 12.9~$\mu$m, while a larger $R_{in}$ = 650~au decreases it 3.8 times at 16.6~$\mu$m compared to the best-fit flux with $R_{in}$ = 550~au. The variation of $R_{out}$ changes SED at longer wavelength: a smaller $R_{out}$ = 600~au causes the maximum flux to decrease 2.9 times at 155.9~$\mu$m, while a larger $R_{out}$ = 1000~au increases it 2.1 times at 95.7~$\mu$m.

\par We model the fluxes from the disc, assuming that it should be tilted at no less than 13\degr as was determined from the analysis of the spectroscopic data (see Section~\ref{spectra}). Our computations indicate that the disc rather have a large inclination angle. The middle-right panel of Fig.~\ref{SEDs_var} shows that if the disc is less tilted toward the observer, the flux would be stronger in the wavelength interval from $\sim$10 to $\sim$200~$\mu$m. In particular, the flux at $\lambda = 29.2~\mu$m increases by a factor of 1.9, when the inclination angle is equal to 20\degr.

\par The change of $q$ (bottom left panel) from 0.75 to 0.5 causes a 35~$\%$ flux decrease at $\lambda = 15$~$\mu$m. We also analysed the influence of the  power law of the surface density $p$ (from 1.5, as we assume in the present work, down to 0 - when $\varSigma_r$ is constant with the distance). As in the case of $q$ it has almost no effect on the SED. This is explained by the fact that the temperature and surface density are described by power laws that cause a small effect at very large distances. A large value of $R_{in}$ and an assumption that the total disc mass remains the same (that we make for the present analysis) lead to a negligible change in the SED for different temperature and surface density distributions.
 
\par We can determine only a maximum value for the disc vertical height at the inner edge within the modelling errors (see Fig. 6, bottom right panel). The disc of larger vertical height would have a significantly stronger flux. If $H_{in}=10^{-2}R_{in}$, the flux from the system would be only 9.4~$\%$ stronger at 160~$\mu$m (the longest wavelength of our flux dataset), while if $H_{in}=10^{-1}R_{in}$, the flux from the system would be almost 2 times stronger at 160~$\mu$m.

\par We can determine only a minimum value of the disc total mass $M_d$ within the modelling errors (see the middle-left panel). The disc of a smaller mass would have a noticeably weaker flux. For example, for $M_d = 2\times10^{-2}M_{\odot}$, the flux from the system becomes 2.8~times smaller at 160~$\mu$m, while for significantly more massive discs (for example $M_d = 2 M_{\odot}$), the flux becomes only 1.5~$\%$ stronger at 160~$\mu$m. Bigger $M_d$ increases the disc total flux at longer wavelengths more noticeably, that is why additional observations are needed for a more precise determination of this parameter.

\subsection{Comparison with other systems}
\label{discussion}

\par
\begin{figure}
\centering
\includegraphics[width=8.0cm, bb =  50   20   710   540, clip=]{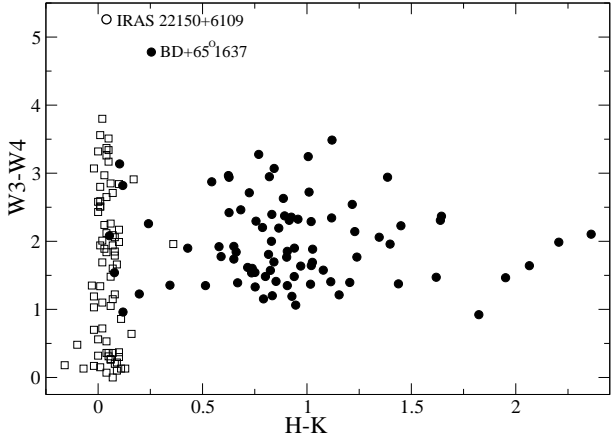} 
\caption{$H-K$ versus $W3-W4$ colour-colour diagramme for Herbig Ae/Be stars from \citet{The94} shown by filled circles
and debris disc candidates from \citet{Liu2014} shown by open squares.
$W3$ and $W4$ are the {\it WISE} magnitudes at 12 and 22 $\mu$m, respectively.}
\label{WISE_diagr}  
\end{figure}

 The infrared excess of IRAS~22150+6109, as mentioned in Section\,\ref{introduction}, is similar to those of young stars which still retain circumstellar material left from the formation phase. It has been shown that such objects lose the near-infrared part of the excess first due to the effect of radiation pressure on dusty particles that move away from the star \citep[e.g.,][]{mir96,mal98}. The pre-main-sequence evolutionary time gets shorter with increasing the stellar mass \citep{Palla93}. Therefore, it is harder to catch a higher-mass star in transition to the main sequence phase. Indeed, there are only a few known Herbig Be stars with almost no near-infrared excess and a strong far-infrared one. The most similar one to IRAS~22150+6109 in terms of the SED is BD+65$^{\circ}$1637 \citep{Patel2016}. These two objects show very different near-to-mid infrared colours from those of other Herbig Ae/Be stars (Fig.\,\ref{WISE_diagr}). At the same time, objects with debris discs, which typically exhibit no emission lines in the optical spectra and have already started the main-sequence evolutionary phase, also occupy a well-defined region on this diagramme. They have already lost the near-infrared excess, while their mid-infrared excess is weaker than that of the two transitional objects. Therefore, the diagramme can be used for finding other transitional objects, whose emission-line spectra are already weak.

\section{Conclusions}
\label{conclusions}

 The new optical spectroscopy and optical multicolor photometry of  the infrared source IRAS~22150+6109 suggests that the object is a B3 star surrounded by a gaseous and dusty disc. The ratio of the observed projected rotational velocity to the critical velocity sets a limit on the tilt angle of the star's rotational axis of $i > 13\degr$. The interstellar features detected in the spectrum are consistent with the star's location at a distance comparable to that of the dark cloud L\,1188. The presence of emission lines supports an earlier suggestion that it is a young star that was based on the detection of the infrared excess. From our SED modelling we found that the disc has been swept away from the central star to a distance of $\sim$550~AU and has a very large outer radius of $\sim$700AU. The disc inclination is $ \geq 60\degr$, and its mass $M_{d}$$> 3 \times 10^{-2} M_{\ast}$. Comparison with other young stars \citep{The94,Liu2014} indicates that IRAS~22150+6109 is one of a very few known Herbig Be stars with an almost no near-infrared excess and a very strong far-infrared one. It is most likely in transition to the main-sequence phase which, as expected, is very short \citep[$\sim10^{4}$ years, e.g.,][]{Palla93} for stars of this mass range. This makes IRAS~22150+6109 a very interesting and important target for future studies.

\section*{Acknowledgements}
The authors thank  Y.~Boehler for providing the dust opacities for SED modelling. ASM acknowledges support from the University of North Carolina at Greensboro College of Arts \& Sciences Advancement Council and the UNCG Department of Physics and Astronomy. ASM and SVZ acknowledge support from DGAPA/PAPIIT Project IN100617. The results are partially based on observations obtained at the Canada-France-Hawaii Telescope (CFHT) which is operated by the National Research Council of Canada, the Institut National des Sciences de l$^{\prime}$Univers of the Centre National de la Recherche Scientifique de France, and the University of Hawaii as well as  upon observations carried out at the Observatorio Astron\'omico Nacional on the Sierra San Pedro M\'artir, Baja California, M\'exico.  The work was carried out within the framework of Project No. BR05236322 and BR05236494, financed by the Ministry of Education and Science of the Republic of Kazakhstan.

\end{document}